\documentclass[showpacs,aps,prd,reprint,superscriptaddress,nofootinbib,longbibliography]{revtex4-2}
\usepackage[colorlinks=true, pdfstartview=FitV,
bookmarks=true, bookmarksnumbered=true, breaklinks]{hyperref}
\usepackage[dvipdfmx]{graphicx}
\usepackage{amsmath,amssymb,bm,color,longtable,mathrsfs,slashed,comment,tikz,enumerate}

\definecolor{darkviolet}{rgb}{0.58, 0.0, 0.83}
\definecolor{electricultramarine}{rgb}{0.25, 0.0, 1.0}
\definecolor{brightpink}{rgb}{1.0, 0.0, 0.5}
\hypersetup{linkcolor=brightpink, citecolor=darkviolet, urlcolor=electricultramarine}

\definecolor{lime}{HTML}{A6CE39}
\DeclareRobustCommand{\orcidicon}{
	\hspace{-3mm}
	\begin{tikzpicture}
	\draw[lime, fill=lime] (0,0) 
	circle [radius=0.16] 
	node[white] {{\fontfamily{qag}\selectfont \tiny ID}};
	\draw[white, fill=white] (-0.0625,0.095) 
	circle [radius=0.007];
	\end{tikzpicture}
	\hspace{-3mm}
}

\foreach \x in {A, ..., Z}{\expandafter\xdef\csname orcid\x\endcsname{\noexpand\href{https://orcid.org/\csname orcidauthor\x\endcsname}
			{\noexpand\orcidicon}}
}

\begin{document}

\title{Anomalous-magnetic-moment-enhanced Casimir effect}

\author{Daisuke Fujii\orcidA{}}
\email[]{daisuke@rcnp.osaka-u.ac.jp}
\affiliation{Advanced Science Research Center, Japan Atomic Energy Agency (JAEA), Tokai, 319-1195, Japan}
\affiliation{Research Center for Nuclear Physics, Osaka University, Ibaraki 567-0048, Japan}

\author{Katsumasa~Nakayama\orcidB{}}
\email[]{katsumasa.nakayama@riken.jp}
\affiliation{RIKEN Center for Computational Science, Kobe, 650-0047, Japan}

\author{Kei~Suzuki\orcidC{}}
\email[]{k.suzuki.2010@th.phys.titech.ac.jp}
\affiliation{Advanced Science Research Center, Japan Atomic Energy Agency (JAEA), Tokai, 319-1195, Japan}

\begin{abstract}
We theoretically investigate the impact of the anomalous magnetic moment (AMM) of Dirac fermions on the fermionic Casimir effect under magnetic fields.
We formulate it as an extension of the well-known Lifshitz formula.
From our formula, we find that the AMM increases the fermionic Casimir energy.
In particular, when the AMM is large enough, the Casimir energy is significantly enhanced by the gapless behavior of the lowest Landau level.
We also quantitatively estimate the Casimir energy from electron, muon, and constituent quark fields under magnetic fields and propose possible phenomena at finite temperature and fermion density.
\end{abstract}

\maketitle

\section{Introduction} \label{Sec:1}

In 1948, Casimir predicted an attractive force working between parallel conducting plates, which is induced by a modification of the zero-point energy of photon fields in quantum electrodynamics (QED) vacuum~\cite{Casimir:1948dh}.
About fifty years later, this effect was experimentally confirmed~\cite{Lamoreaux:1996wh,Bressi:2002fr} (see Refs.~\cite{Plunien:1986ca,Mostepanenko:1988bs,Bordag:2001qi,Milton:2001yy,Klimchitskaya:2009cw,Woods:2015pla,Gong:2020ttb,Lu:2021jvu} for reviews).
Nowadays, related phenomena are expected to be applied in the field of engineering, such as nanophotonics~\cite{Gong:2020ttb}, where a significant issue is {\it how to control the Casimir effect through external environments}, such as temperature, density, and external fields.
In this paper, we focus on magnetic fields.

The conventional Casimir effect (originating from the photon field) is hardly affected by magnetic fields because photons do not directly couple to magnetic fields~\cite{Robaschik:1986vj},\footnote{As another situation related to magnetic fields, one can consider the magnetic property of boundary conditions (e.g., \cite{Bruno:2002,Metalidis:2002,Grushin:2010qoi}), but this is different from our purpose in our current study.} whereas charged particles are directly affected.
For example, Casimir effects originating from quantum fields under a magnetic field have been widely studied for charged scalar fields~\cite{Cougo-Pinto:1998jun,Cougo-Pinto:1998mdw,Cougo-Pinto:1998fpo,Cougo-Pinto:1998zge,Elizalde:2002kb,Ostrowski:2005rm,Erdas:2013jga,Erdas:2013dha,Sitenko:2014wwa,Sitenko:2015eza,Erdas:2015yac,Erdas:2020ilo,Haridev:2021jwi, Erdas:2021xvv,Erdas:2024bxd,Droguett:2025frq,Erdas:2025gbv} as well as the Dirac field~\cite{Cougo-Pinto:1998jwo,Cougo-Pinto:2001kks,Elizalde:2002kb,Ostrowski:2005rm,Miltao:2008zza,Sitenko:2014kza,Sitenko:2015eza,Sitenko:2015wzd,Nakayama:2022fvh,Rohim:2023tmy,Erdas:2023wzy,Flachi:2024ztd,Fujii:2024woy}.
As more realistic setup, one can consider fermionic Casimir effects inside thin films of Dirac (or Weyl) semimetals~\cite{Nakayama:2022fvh} and thin quark matter~\cite{Fujii:2024woy}.
In these systems, the Casimir effect can be observed as a change of thermodynamic quantities (e.g., pressure, specific heat, and magnetization) as a function of thickness, and one could control the Casimir effect by changing the strength of magnetic fields.

In the study of the Casimir effect under a magnetic field, a missing piece is the contribution of the anomalous magnetic moment (AMM) of charged fields.
A representative example of AMM is a shift of the electron $g$ factor from $2$, which was calculated by Schwinger in 1948~\cite{Schwinger:1948iu,Schwinger:1949ra}.  
The role of AMM on the Casimir effect has never been studied in prior studies~\cite{Cougo-Pinto:1998jun,Cougo-Pinto:1998mdw,Cougo-Pinto:1998fpo,Cougo-Pinto:1998zge,Ostrowski:2005rm,Erdas:2013jga,Erdas:2013dha,Sitenko:2014wwa,Sitenko:2015eza,Erdas:2015yac,Erdas:2020ilo,Haridev:2021jwi, Erdas:2021xvv,Erdas:2024bxd,Droguett:2025frq,Erdas:2025gbv,Cougo-Pinto:1998jwo,Cougo-Pinto:2001kks,Elizalde:2002kb,Miltao:2008zza,Sitenko:2014kza,Sitenko:2015wzd,Nakayama:2022fvh,Rohim:2023tmy,Erdas:2023wzy,Flachi:2024ztd,Fujii:2024woy}.
As a naive expectation, because a nonzero AMM modifies the eigenvalue spectrum (i.e., Landau levels) of charged particles, the AMM can also affect the property of the Casimir effect.

In this work, we investigate possible contributions of AMMs on the Casimir effect.
In QED, the AMM of electrons in a vacuum or in a weak magnetic field is usually tiny, but a strong magnetic field may induce a dynamical Zeeman effect (or dynamical AMM)~\cite{Ferrer:2008dy,Ferrer:2009nq}.
Also, in the low-energy region of quantum chromodynamics (QCD), up, down, and strange quarks have dynamical masses due to the spontaneous chiral symmetry breaking, and a part of their magnetic moments can be regarded as an AMM in the corresponding Dirac equation~\cite{Singh:1985sg,Bicudo:1998qb,Chang:2010hb}.
In the context of the QCD phase diagram in magnetic fields, one can construct effective models based on constituent-quark degrees of freedom affected by such an AMM~\cite{Fayazbakhsh:2014mca,Chaudhuri:2019lbw,Ghosh:2020xwp,Xu:2020yag,Mei:2020jzn,Aguirre:2020tiy,Wen:2021mgm,Aguirre:2021ljk,Chaudhuri:2021skc,Wang:2021dcy,Chaudhuri:2021lui,Lin:2022ied,Kawaguchi:2022dbq,Mao:2022dqn,Sheng:2022ssp,Chaudhuri:2022oru,He:2022inw,Qiu:2023kwv,Tavares:2023oln,He:2024gnh,Mondal:2024eaw,Kawaguchi:2024edu,Yang:2025zcs} (or Lorentz-tensor condensates~\cite{Ferrer:2013noa,Mao:2018jdo,Lin:2022ied,Qiu:2023kwv,Bao:2024glw}, which leads to a similar but slightly different effect).
Also in such situations, an AMM should play an important role in the magnetic response of the Casimir effect.

This paper is organized as follows.
In Sec.~\ref{Sec:2}, we formulate the Casimir effect from Dirac fields with AMMs.
In Sec.~\ref{Sec:3}, the typical behavior of the Casimir effect and numerical results are discussed.
In Sec.~\ref{Sec:4}, related topics are discussed. 
Section~\ref{Sec:5} is devoted to the conclusion.

\section{Formulation} \label{Sec:2}

\subsection{Dirac field with AMM}

In this work, we consider Dirac fields, $\psi$ and $\bar{\psi} \equiv \psi^\dagger \gamma_0$, with a mass $m$ and an electric charge $q$.
As specific forms of electric charges, $q_{e/\mu}=-e$ for the electron or the muon, $q_u=\frac{2}{3}e$ for the up quark, and $q_{d/s}=-\frac{1}{3}e$ for the down and strange quarks, where $e>0$ is the elementary charge.

The Lagrangian of the Dirac field is
\begin{align}
\mathcal{L} = \bar{\psi} (i\gamma^\mu D_\mu -m +\frac{\kappa q}{2} \sigma^{\mu\nu} F_{\mu\nu} )\psi,
\end{align}
where $D_\mu \equiv \partial_\mu + iqA_\mu$ and $F_{\mu\nu} = \partial_\mu A_\nu -\partial_\nu A_\mu$ are the covariant derivative and the field strength tensor with an external U(1) gauge field $A_\mu$, respectively, and $\sigma^{\mu\nu} = \frac{i}{2} (\gamma^\mu \gamma^\nu -\gamma^\nu \gamma^\mu)$.
The AMM is characterized by a parameter $\kappa$.
When a magnetic field is applied in the $z$-axis, $\vec{B} =(0,0,B)$, and the Landau gauge $A_\mu=(0,0,Bx,0)$ is chosen,
\begin{align}
\mathcal{L} = \bar{\psi} (i\gamma^\mu D_\mu -m + \kappa q \sigma^{12} B )\psi.
\end{align}

By diagonalizing the Hamiltonian in momentum space, we obtain the dispersion relations~\cite{Ternov:1966},
\begin{align}
\omega_{l, s} &= E_{l, s}, \\
\tilde{\omega}_{l, s} &= -E_{l, s},
\end{align}
where
\begin{align}
E_{l, s} &= \sqrt{k_z^2 +  \left( \sqrt{m^2 + (2l+1- s \zeta) |qB|} - s\kappa qB \right)^2}, \label{eq:E_ls}
\end{align}
which is labeled by the quantum number of discrete levels $l=0,1,2, \dots$, the spin $s=\pm 1$, and $\zeta=\mathrm{sgn}(qB)$.
When we set $\zeta=+1$, the lowest modes are the set of $\omega_{l=0,+}$ and $\tilde{\omega}_{l=0,+}$.
We call this set the {\it lowest Landau levels (LLLs)}, and the other levels {\it higher Landau levels (HLLs)}.

\begin{figure}[tb!]
    \centering
    \begin{minipage}[t]{1.0\columnwidth}
   \includegraphics[clip,width=1.0\columnwidth]{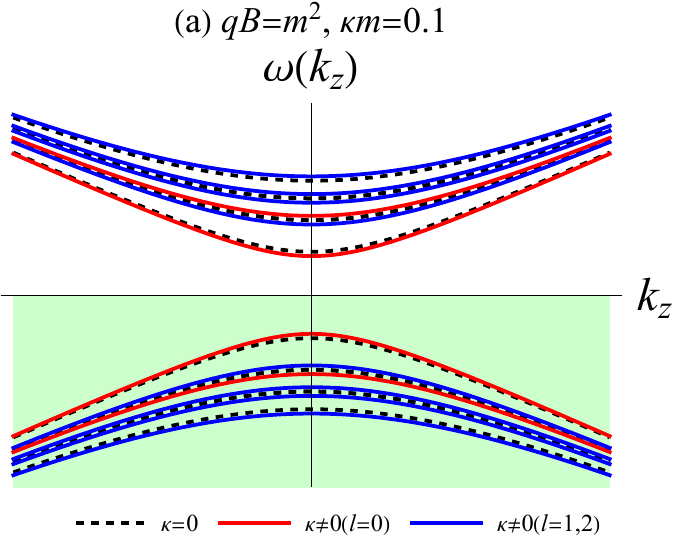}
    \end{minipage}
    \begin{minipage}[t]{1.0\columnwidth}
   \includegraphics[clip,width=1.0\columnwidth]{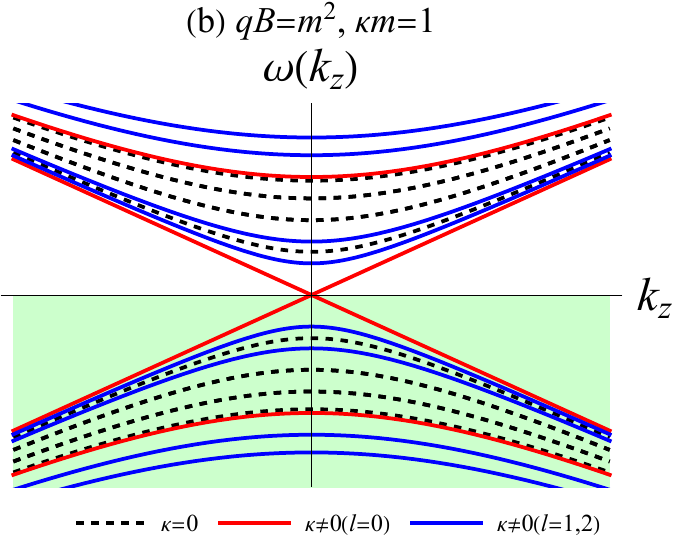}
    \end{minipage}
    \begin{minipage}[t]{1.0\columnwidth}
   \includegraphics[clip,width=1.0\columnwidth]{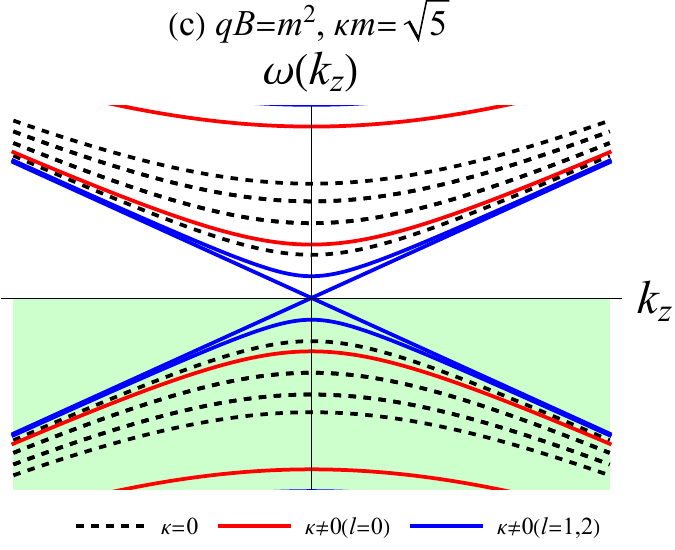}
    \end{minipage}
\caption{
Examples of dispersion relations in massive Dirac fields in a fixed magnetic field at (a) a small $\kappa$, (b) the lowest critical AMM $\kappa^{{\rm cri}(0)}$, and (c) the first critical AMM $\kappa^{{\rm cri}(1)}$.
}
\label{fig:disp}
\end{figure}

Typical examples of the $\kappa$-dependence of dispersion relations in a fixed $qB=m^2$ are shown in Fig.~\ref{fig:disp}.
\begin{enumerate}[(i)]
\item At $\kappa=0$, the LLLs are spin-projected, whereas the HLLs are doubly degenerate between, e.g., $\omega_{l=0,-}$ and $\omega_{l=1,+}$, which is a well-known feature of the Dirac field under magnetic fields~\cite{Rabi:1928,Johnson:1949} (see the black dashed lines in Fig.~\ref{fig:disp}).
\item At $\kappa \neq 0$, we can see that a nonzero AMM splits the spin degeneracy of the HLLs, and the gap of the LLL, characterized by the mass $m$, gets smaller [see Fig.~\hyperref[fig:disp]{\ref*{fig:disp}(a)}].
\item At a critical AMM $\kappa^{{\rm cri}(0)}=m/qB$, which we call the {\it lowest critical AMM}, the LLLs becomes a linear dispersion relation due to the cancellation of the mass and the AMM~\cite{O'Connell:1968,Chiu:1968zz} [see the red solid lines in Fig.~\hyperref[fig:disp]{\ref*{fig:disp}(b)}].
\item At a larger $\kappa > m/qB$, the gap of the LLLs become opened again, whereas the gap of the first LLs (i.e., $l=1$) decreases with increasing $B$.
\item At $\kappa^{{\rm cri}(1)}=\sqrt{m^2+2|qB|}/qB$, which we call the {\it first critical AMM}, the gap of the first LLs vanishes, and the linear dispersion relations appears [see the blue solid lines in Fig.~\hyperref[fig:disp]{\ref*{fig:disp}(c)}].
\item At a larger $\kappa>\sqrt{m^2+2|qB|}/qB$, the first LLs become gapped again.
\item This series of behaviors holds for any $l$, where each critical AMM is $\kappa^{{\rm cri}(l)} = \sqrt{m^2+2l|qB|}/qB$.
\item We note that this sequence occurs also when we vary the magnetic field at a fixed $\kappa$~\cite{Chiu:1968zz}, where the critical magnetic field for the LLLs is $qB^{{\rm cri}(0)}=m/\kappa$.
In the following analysis, we will focus on this magnetic field as a typical parameter, and we simply denote it by $qB^{{\rm cri}} \equiv qB^{{\rm cri}(0)}$.
\end{enumerate}

\subsection{Zero-point energy}
Using the dispersion relations, the thermodynamic potential $\Omega \equiv -\frac{T}{V} \ln Z$ (per unit volume $V=L_xL_yL_z$ at a temperature $T=1/\beta$ with the distribution function $Z$) is
\begin{align}
\Omega =& -\frac{|q B|}{2\pi}\int\frac{dk_z}{2\pi} \sum_{l,s} \left[\frac{1}{2}|\omega_{l,s}|+\frac{1}{2}|\tilde{\omega}_{l,s}| \right. \notag \\
& \left. +\frac{1}{\beta} \ln\left\{\left(1+e^{-\beta|\omega_{l,s}|}\right)\left(1+e^{-\beta|\tilde{\omega}_{l,s}|}\right)\right\} \right], \label{eq:omega}
\end{align}
where the overall minus sign is a property of the fermions. 
The factor of $|qB|/2\pi$ is called the Landau degeneracy factor, which is obtained from the momentum integral $\int dk_xdk_y/(2\pi)^2$ at $B=0$. 
By taking the zero-temperature limit of Eq.~(\ref{eq:omega}), we obtain the zero-point energy of the system,
\begin{align}
    \Omega(T\rightarrow0) &= - \frac{|q B|}{2\pi}\int\frac{dk_z}{2\pi} \sum_{l,s} \left(\frac{1}{2}|\omega_{l,s}|+\frac{1}{2}|\tilde{\omega}_{l,s}|\right) \notag \\
&\equiv \frac{E_0^{\rm int}}{L_z}. \label{E0int} 
\end{align}
Note that we defined $E_0^{\rm int}$ as the zero-point energy per unit area $L_xL_y$ (i.e., an energy surface density), and its mass dimension is $3$.

\subsection{Casimir energy}
To discuss the Casimir energy, we consider the zero-point energy in a finite volume.
In this work, we impose the periodic boundary conditions (PBCs) at $z=0$ and $z=L_z$ in the $z$ direction,\footnote{This setup is easy because $k_z$ is a conserved quantity in systems under a magnetic field, whereas the case with boundary conditions imposed in the $x$ or $y$ direction is analytically difficult.} where the spatial momentum is discretized as $k_z= 2n\pi/L_z$ ($n=0,\pm 1,\pm 2, \dots$).
Note that the Casimir effect under the PBCs is the simplest setup that can be studied by numerical simulations of the lattice field theory, such as lattice QED and QCD.
In the future, such simulations enable us to quantitatively investigate the Casimir effect modified by nonperturbative phenomena, such as dynamical AMMs and chiral symmetry breaking.
Our current study provides theoretical grounds for guiding future numerical simulations.\footnote{We comment on the physical meaning of the Casimir effect under PBCs.
Intuitively, the internal pressure of the material compactified by PBCs is affected by the finite-volume effect.
As a result, the compactified direction can be shrunk or expanded (if the volume of systems is changeable).
This phenomenon is analogous to the attractive (or repulsive) force between conventional parallel plates.
In the real world, such a geometry can be realized as one-dimensional closed strings or winding directions in two-dimensional nanotubes (e.g., carbon nanotubes) and three-dimensional solid tori.
For example, in winding materials composed of fermions, the length of the winding direction can be shrunk or expanded by the fermionic Casimir pressure, compared to the case with infinite length.}

Then, using the replacements of $L_z \int dk_z/2\pi \to \sum_n$, $\omega_{l,s} \to \omega_{l,s,n}$, and $\tilde{\omega}_{l,s} \to \tilde{\omega}_{l,s,n}$ in Eq.~(\ref{E0int}), the zero-point energy (per unit area) at finite $L_z $ is
\begin{align}
E_0^{\rm sum} \equiv -\frac{|qB|}{2\pi} \sum_{l,s} \sum_{n=-\infty}^\infty
\left( \frac{1}{2}|\omega_{l,s,n}|+\frac{1}{2}|\tilde{\omega}_{l,s,n}| \right). \label{E0sum}
\end{align}
This infinite sum (\ref{E0sum}) has an ultraviolet divergence.
To get the finite Casimir energy, we need a regularization scheme.

The Lifshitz formula is a powerful tool that was first derived for calculating the conventional photonic Casimir energy by Lifshitz~\cite{Lifshitz:1956zz}.
The corresponding formula for the Dirac field in a nonzero magnetic field was used in Ref.~\cite{Fujii:2024woy}, which is equivalent to the result derived from other regularization schemes (e.g., the proper-time regularization~\cite{Cougo-Pinto:1998jwo,Cougo-Pinto:2001kks}, the Abel-Plana formula~\cite{Ostrowski:2005rm}, the zeta-function regularization~\cite{Miltao:2008zza}, and the lattice regularization~\cite{Fujii:2024woy}).

In this work, we apply the Lifshitz formula with an AMM,
\begin{align}
    E_\mathrm{Cas} =&-2\int_{-\infty}^\infty \frac{d\xi}{2\pi} \sum_{l,s} \frac{|qB|}{2\pi} \ln \left[1-e^{-L_z \tilde{k}^{[l,s]}_z} \right], \label{Lifshitz} \\ 
    \tilde{k}^{[l,s]}_z =& \sqrt{\left(\sqrt{ m^2 + \left(2l+1-s \zeta \right)|qB|} - s \kappa q B \right)^2 +\xi^2}. \notag 
\end{align}
This is the main formula in the current work.
The overall factor $-2$ is the minus sign of the fermion zero-point energy and the factor due to the PBC.\footnote{It is also straightforward to derive the Lifshitz formula under the antiperiodic boundary conditions (e.g., see Ref.~\cite{Fujii:2024ixq}).
In addition, for the MIT bag boundary conditions~\cite{Chodos:1974je}, the Casimir effect from Dirac fields (without AMM) under magnetic fields was analyzed by using a subtraction scheme with the Mittag-Leffler's theorem~\cite{Elizalde:2002kb}, the Abel-Plana formula~\cite{Sitenko:2014kza,Sitenko:2015eza,Sitenko:2015wzd,Rohim:2023tmy}, and the zeta-function regularization~\cite{Erdas:2023wzy}.
Using these approaches, the analysis with the AMM is possible.
Note that the Lifshitz formula for gapless dispersions under the MIT bag boundary conditions is given in Ref.~\cite{Fujii:2024ixq}.}
The integral variable $\xi$ is the imaginary part of the imaginary energy, where the integral over $\xi$ covers the contributions of the particle and antiparticle degrees of freedom.
At $\kappa=0$, this formula is reduced to the formula used in Ref.~\cite{Fujii:2024woy}.\footnote{Precisely speaking, we need to substitute $b=\mu=0$ and remove $N_c \sum_{q_f}$ in Eq.~(11) of Ref.~\cite{Fujii:2024woy}.
Note that there are two ways to assign indices of the Landau levels: the two modes, $\omega_{l=0,+}$ and  $\omega_{l=0,-}$, assigned in this work are written as $\omega_{l=0}$ and either of $\omega_{l=1,\pm}$ in Ref.~\cite{Fujii:2024woy}, respectively.}

\section{Results} \label{Sec:3}

\subsection{Typical behaviors}

\begin{figure}[tb!]
    \centering
    \begin{minipage}[t]{1.0\columnwidth}
    \includegraphics[clip,width=1.0\columnwidth]{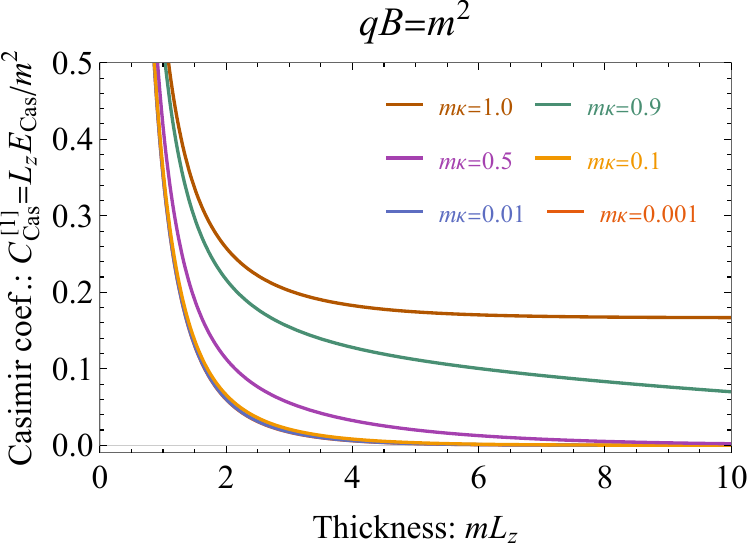}
    \end{minipage}
\caption{
Casimir coefficient $C_\mathrm{Cas}^{[1]} \equiv L_z E_\mathrm{Cas}$ without ($\kappa=0$) and with AMM ($\kappa \neq 0$).
}
\label{fig:typ1}
\end{figure}

Figure~\ref{fig:typ1} shows the $\kappa$ dependence ($\kappa m =0, 0.01, 0.1, 0.5, 1$) of the Casimir energy in a fixed magnetic field $qB=m^2$.
Here, we plot a Casimir coefficient
\begin{align}
C_\mathrm{Cas}^{[1]} \equiv L_z E_\mathrm{Cas},
\end{align}
where the superscript $[1]$ denotes the power of $L_z$ in the right-hand side.
This quantity is useful for focusing on the {\it $1/L_z$ scaling behavior}.
In the longer-$L_z$ region, the Casimir energy is dominated by the LLLs, because they are the lightest among all LLs within the current parameters.
From Fig.~\ref{fig:typ1}, we find that a nonzero $\kappa$ leads to an enhancement of the Casimir energy.
This increase is caused by the reduction of the energy gaps of the LLLs in the dispersion relations.
In particular, at the critical AMM ($m\kappa = 1$), the Casimir energy $E_\mathrm{Cas}$ has a clear $1/L_z$ scaling, which is a typical feature for gapless dispersions.
This is one of our main findings in this work: when the AMM is large enough, the LLLs become effectively gapless, which results in a significant enhancement of the Casimir energy.
We note that the gapless dispersion also occurs even when the magnetic field is strong enough at a fixed AMM~\cite{O'Connell:1968,Chiu:1968zz}, where the critical magnetic field is $qB^\mathrm{cri} =m/\kappa$.
Therefore, even if the AMM is small, a strong magnetic field can significantly enhance the Casimir effect.

On the other hand, when the AMM is smaller compared to the mass ($m\kappa < 1$), the long-$L_z$ behavior is strongly suppressed, which is a nonzero-mass effect characterized by $ m\neq0 $ and a well-known behavior for massive fields~\cite{Hays:1979bc,Mamaev:1980jn,Ambjorn:1981xw}.

In the shorter-$L_z$ region of Fig.~\ref{fig:typ1}, the Casimir energy increases, which indicates an enhancement stronger than the $1/L_z$ scaling.
This is because the higher LLs (as well as the LLLs) contribute to the Casimir effect.
We find that, also in this region, the Casimir energy is enhanced by the AMM.

\begin{figure}[t!]
    \centering
    \begin{minipage}[t]{1.0\columnwidth}
    \includegraphics[clip,width=1.0\columnwidth]{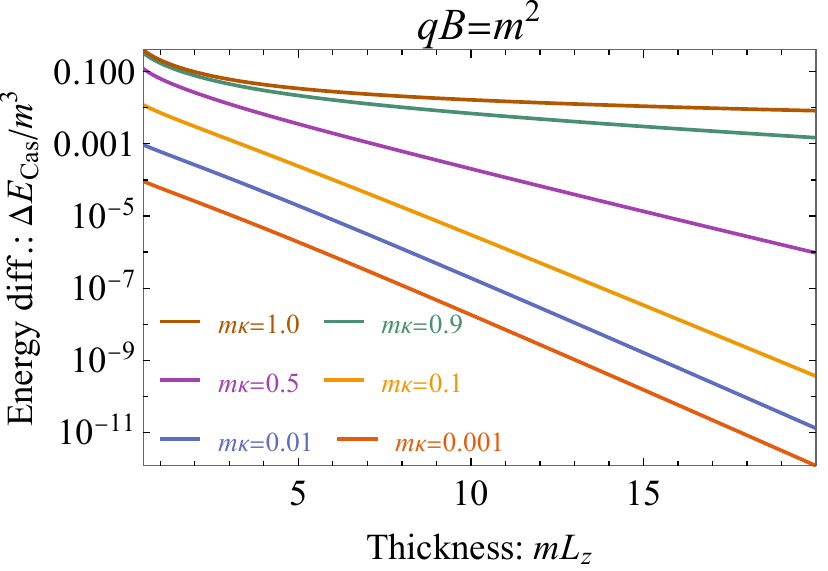}
    \end{minipage}
\caption{
Semi-log plot of Casimir energy enhancement $\Delta E_\mathrm{Cas}$ due to $\kappa \neq 0$.
}
\label{fig:typ_log}
\end{figure}

Next, to examine the significance of the AMM effect, we define the Casimir energy difference at $\kappa \neq 0$ and $\kappa=0$ as
\begin{align}
\Delta E_\mathrm{Cas} \equiv E_\mathrm{Cas}(\kappa) - E_\mathrm{Cas} (\kappa=0).
\end{align}
A semi-log plot is useful for examining {\it exponential scaling behavior} in the long-$L_z$ region.
In Fig.~\ref{fig:typ_log}, we find that $\Delta E_\mathrm{Cas}$ behaves linearly in the long-$L_z$ region $mL_z>1$.
This indicates that $\Delta E_\mathrm{Cas}$ decays exponentially in this region.
At larger $\kappa$ ($m\kappa=0.5$), the slope of linear behavior is determined by both $m$ and $\kappa$: $\Delta E_\mathrm{Cas} \propto e^{-(m-\kappa |qB|)L_z}$.
At smaller $\kappa$ ($m\kappa=0.001,0.01,0.1$), the slope is mostly determined by $m$, and the $\kappa$ dependence is suppressed: the exponential scaling in this region is $\Delta E_\mathrm{Cas} \propto e^{-mL_z}$.
These scaling laws arise from the LLL dominance in the long $L_z$ region.\footnote{
An expression of the Casimir energy for the LLLs, which is equivalent to the $l=0$ term in Eq.~(\ref{Lifshitz}), is given as
\begin{align}
E_\mathrm{Cas,LLL} = \frac{|qB|}{2\pi} \times \frac{2M}{\pi} \sum_{j=1}^\infty \frac{K_1[jML_z]}{j},
\end{align}
where $K_1$ is the modified Bessel function, and $M=m-\kappa |qB|$ within $M>0$.
By focusing on the $j=1$ term and using $K_1(x) \approx \sqrt{\frac{\pi}{2x}}e^{-x}$ for $x\gg 1$, we can get an exponential-type function.
}
Note that near the critical AMM ($m\kappa \sim 1$), $\Delta E_\mathrm{Cas} \propto 1/L_z$, which is not exponential scaling.

\begin{figure}[tb!]
    \centering
    \begin{minipage}[t]{1.0\columnwidth}
    \includegraphics[clip,width=1.0\columnwidth]{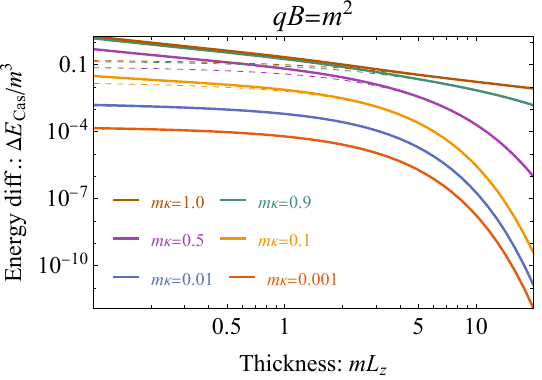}
    \end{minipage}
\caption{
Log-log plot of Casimir energy enhancement $\Delta E_\mathrm{Cas}$ due to $\kappa \neq 0$.
Solid lines: the results with the LLLs and HLLs.
Dashed lines: with only LLLs.
}
\label{fig:typ_loglog}
\end{figure}

A log-log plot is useful for focusing on the short-$L_z$ region and also examining the {\it power-law scaling behavior} of a physical quantity.
In Fig.~\ref{fig:typ_loglog}, we find that $\Delta E_\mathrm{Cas}$ behaves linearly at a short $L_z$ (now, $mL_z \lesssim 2$), which is the power-law scaling.
In addition, in this region, we can examine the contribution of each LL.
The dashed lines in Fig.~\ref{fig:typ_loglog} represent the results with only the LLLs.
We find that, at larger $\kappa$, the HLLs (as well as the LLLs) contribute to the enhancement of the Casimir energy.
On the other hand, at smaller $\kappa$, the LLLs are still dominant.

\subsection{Electrons}
Next, we discuss the Casimir energy from the electron field in the QED vacuum.
In the zero magnetic field, the intrinsic AMM is a well-known value $a \equiv (g-2)/2 = \alpha/2\pi$ in the one-loop level, which was derived by Schwinger~\cite{Schwinger:1948iu,Schwinger:1949ra}.
A nonzero magnetic field induces an additional AMM.
Within the perturbation theory in the first order of $\alpha$, the induced AMM tends to weaken the intrinsic AMM $\alpha/2\pi$ in both the weak and strong magnetic fields~\cite{Gupta:1949,Demeur:1953,Newton:1954zz,Ternov:1969,Ritus:1969,Jancovici:1969exc,Tsai:1973jex,Baier:1975uj}.
However, it can be nonperturbatively enhanced by a strong magnetic field~\cite{Ferrer:2008dy,Ferrer:2009nq}.
In this section, for simplicity, we input $\alpha/2\pi$.

By substituting the experimental values~\cite{ParticleDataGroup:2024cfk}, $a_e=0.00115965218062$ and $m_e = 0.51099895000$ MeV, into $\kappa \equiv a/2m$, we obtain
\begin{align}
\kappa_{e} = 0.00113469 \ \mathrm{MeV}^{-1}.
\end{align}
By inputting these parameters into Eq.~(\ref{Lifshitz}), we can estimate the Casimir energy.\footnote{The magnetic-field dependence of the parameters, $m_e$ and $\kappa_e$, is omitted to simplify the interpretation of the results.}

\begin{figure}[tb!]
    \centering
    \begin{minipage}[t]{1.0\columnwidth}
    \includegraphics[clip,width=1.0\columnwidth]{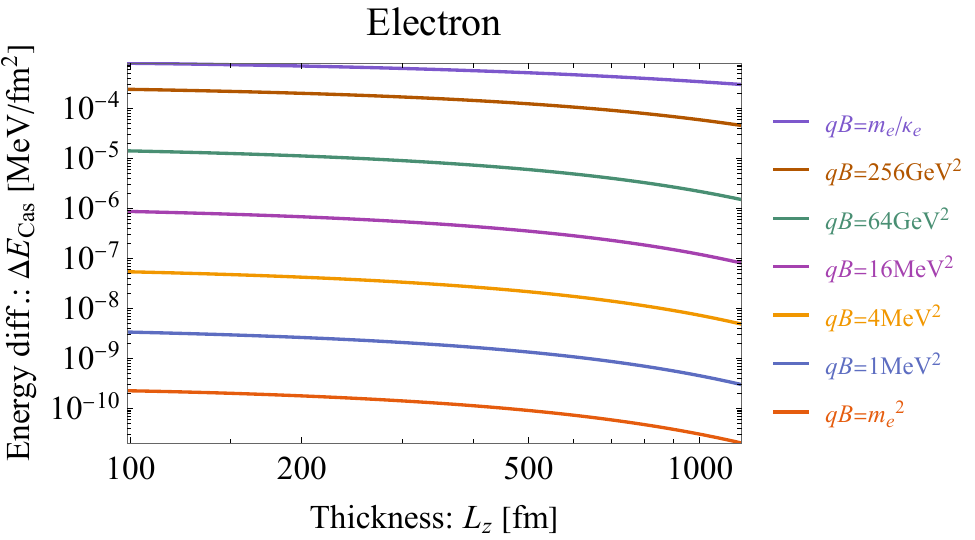}
    \end{minipage}
\caption{Casimir energy enhancement $\Delta E_\mathrm{Cas}$ for electrons.}
\label{fig:electron}
\end{figure}

The numerical results are shown in Fig.~\ref{fig:electron}, where we show a wide range of magnetic fields from $qB=m_e^2 \sim 0.25$ MeV$^2$ (i.e., the Schwinger limit) to the critical magnetic field $qB^\mathrm{cri}=m_e/\kappa_e \sim 450 \mathrm{MeV}^2$.
From this figure, we find that the Casimir energy is enhanced with increasing the magnetic field.
For example, at $qB=m_e^2$, the enhancement is $\Delta E_\mathrm{Cas} \sim 2.3\times 10^{-10} \, \mathrm{MeV/fm}^2$ at $L_z=100$ fm.
At the critical field $ qB^\mathrm{cri}$, $\Delta E_\mathrm{Cas} \sim 7.9\times 10^{-4} \, \mathrm{MeV/fm}^2$ at $L_z=100$ fm.

\subsection{Muons}
The case of muons is similar to that of electrons.
By substituting the experimental values~\cite{ParticleDataGroup:2024cfk}, $a_\mu=0.00116592059$ and $m_\mu = 105.6583755$ MeV, into $\kappa \equiv a/2m$, we obtain
\begin{align}
\kappa_{\mu} = 5.51741 \times 10^{-6} \ \mathrm{MeV}^{-1}.
\end{align}

\begin{figure}[tb!]
    \centering
    \begin{minipage}[t]{1.0\columnwidth}
    \includegraphics[clip,width=1.0\columnwidth]{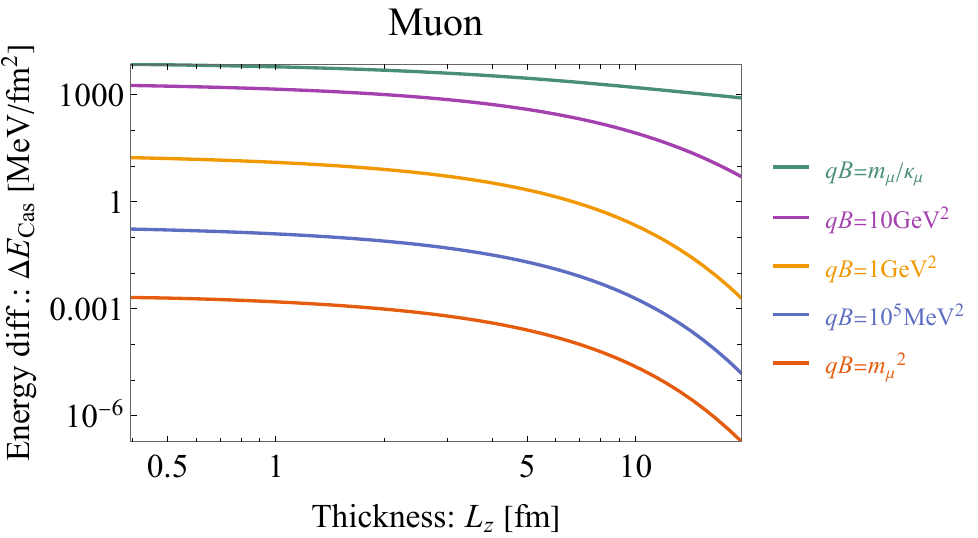}
    \end{minipage}
\caption{Casimir energy enhancement $\Delta E_\mathrm{Cas}$ for muons.}
\label{fig:muon}
\end{figure}

The results are shown in Fig.~\ref{fig:muon}, where we show a wide range of magnetic fields from $qB=m_\mu^2 \sim 1.1\times 10^{4}$ MeV$^2$ to the critical magnetic field $qB^\mathrm{cri}=m_\mu/\kappa_\mu \sim 19 \,\mathrm{GeV}^2$.
We find that the enhancement of the Casimir energy of muons is smaller than that of electrons at the same magnetic field.
Therefore, a significant enhancement is induced by stronger magnetic fields.
For example, at $qB= m_\mu^2$, the enhancement is $\Delta E_\mathrm{Cas} \sim 1.57\times10^{-3} \, \mathrm{MeV/fm}^2$ at $L_z=1$ fm.
At the critical field $ qB^\mathrm{cri}$, $\Delta E_\mathrm{Cas} \sim 6100 \, \mathrm{MeV/fm}^2$ at $L_z=1$ fm.

\subsection{Constituent quarks} \label{subsec:quark}
Contrary to the AMMs of electrons and muons, the AMMs of constituent quarks are nontrivial because they are not gauge-invariant.
However, it is useful in an effective quark model analysis, such as the study of the phase diagram in the Nambu--Jona-Lasinio (NJL) model~\cite{Nambu:1961tp,Nambu:1961fr}.
Therefore, at least our following analysis will be useful for the investigation of the Casimir effect originating from constituent quark fields as degrees of freedom in an effective quark model.
In addition, the AMMs of constituent quarks are related to the magnetic moments of hadrons, and such an AMM-induced effect is interesting also in the context of the hadronic Casimir effect.\footnote{In addition, the Casimir effect in the deconfined phase of QCD is induced by not only quark fields but also gluon fields.
Lattice numerical simulations are powerful for investigating such a gluonic Casimir effect~\cite{Chernodub:2018pmt,Chernodub:2018aix,Kitazawa:2019otp,Chernodub:2023dok}.}

In studies of the phase diagram in the NJL model ~\cite{Fayazbakhsh:2014mca,Chaudhuri:2019lbw,Ghosh:2020xwp,Xu:2020yag,Mei:2020jzn,Aguirre:2020tiy,Wen:2021mgm,Aguirre:2021ljk,Chaudhuri:2021skc,Wang:2021dcy,Chaudhuri:2021lui,Lin:2022ied,Kawaguchi:2022dbq,Mao:2022dqn,Sheng:2022ssp,Chaudhuri:2022oru,He:2022inw,Qiu:2023kwv,Tavares:2023oln,He:2024gnh,Mondal:2024eaw,Kawaguchi:2024edu,Yang:2025zcs}, the quark AMM can be introduced in several approaches: (i) a constant AMM~\cite{Fayazbakhsh:2014mca,Chaudhuri:2019lbw,Ghosh:2020xwp,Xu:2020yag,Mei:2020jzn,Aguirre:2020tiy,Wen:2021mgm,Aguirre:2021ljk,Chaudhuri:2021skc,Chaudhuri:2021lui,Kawaguchi:2022dbq,Mao:2022dqn,Sheng:2022ssp,Chaudhuri:2022oru,Qiu:2023kwv,Tavares:2023oln,Mondal:2024eaw,Yang:2025zcs}, (ii) an AMM proportional to the chiral condensate (i.e., $\kappa \propto \sigma$ with the $\sigma$ mean field)~\cite{Xu:2020yag,Wang:2021dcy,Lin:2022ied,Kawaguchi:2022dbq}, or (iii) an AMM proportional to the square of the chiral condensate (i.e., $\kappa \propto \sigma^2$)~\cite{Lin:2022ied,Kawaguchi:2022dbq,He:2022inw,He:2024gnh,Kawaguchi:2024edu}.
Such assumptions enable systematic investigations of how the phase diagram responds to different treatments of the AMM.

In this work, under the scenario of approach (i), we adopt an estimate based on the framework of the constituent quark model, proposed in Refs.~\cite{Bicudo:1998qb,Fayazbakhsh:2014mca}.
By inputting the experimental values of baryon masses and magnetic moments, we obtain (see Appendix~\ref{App:quarkAMM}),
\begin{align}
\kappa_u &= 0.00952827 \, \mathrm{GeV}^{-1}, \\
\kappa_d &=  0.0831157 \, \mathrm{GeV}^{-1}.
\end{align}
By substituting these AMM parameters into the gap equation in the NJL model, we can evaluate the constituent quark masses modified by an external magnetic field (see Appendix~\ref{App:NJL}).
By inputting the constituent quark masses and the AMM parameters into Eq.~(\ref{Lifshitz}) and multiplying the factor of the number of colors $N_c=3$, we can calculate the Casimir energy for the constituent $u$ and $d$ quarks.\footnote{As a more advanced analysis, it is straightforward to implement the $L_z$ dependence in the gap equation, while we omit it to simplify the interpretation of the results.}

\begin{figure}[tb!]
    \centering
    \begin{minipage}[t]{1.0\columnwidth}
    \includegraphics[clip,width=1.0\columnwidth]{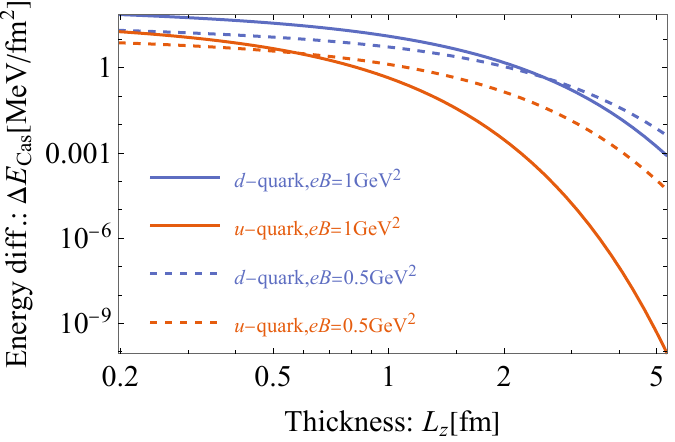}
    \end{minipage}
\caption{Casimir energy enhancement $\Delta E_\mathrm{Cas}$ for constituent $u,d$ quarks.}
\label{fig:quark}
\end{figure}

In Fig.~\ref{fig:quark}, we show the results at $eB=0.5$ and $1\,\mathrm{GeV}^2$.
We find that the enhancement of the Casimir energy for the $d$ quark is larger than for the $u$ quark.
This is because the AMM of the $d$ quark is larger than that of the $u$ quark.
Quantitatively, the energy scale of Casimir energy is similar to that of muons, shown in the previous section.
This is because the effective masses of constituent quarks are similar to the muon mass $m_\mu\sim 100$ MeV. 

In addition, we note that, in the long-$L_z$ region, the Casimir energy (and its enhancement due to the AMM) at $eB =1 \, \mathrm{GeV}^2$ is smaller than the result at $eB= 0.5 \, \mathrm{GeV}^2$.
This is because larger magnetic fields induce heavier dynamical masses.
Thus, the magnetic field does not necessarily enhance the Casimir energy, and a dynamical mass generation can suppress the Casimir energy.

\subsection{At finite temperature}
Next, we consider the situation at finite temperature $T$.
In Eq.~(\ref{Lifshitz}), using the replacement of $\int_{-\infty}^{\infty} \frac{d\xi}{2\pi} f(\xi) \to T \sum_{\mathcal{N}=-\infty}^\infty f(\xi_\mathcal{N})$, where $\xi_\mathcal{N}=(2\mathcal{N}+1)\pi T$ is the fermionic Matsubara frequency, we can obtain the corresponding Lifshitz formula,
\begin{align}
E_\mathrm{Cas} =&-2T \sum_{{\mathcal{N}}=-\infty}^\infty \sum_{l,s} \frac{|qB|}{2\pi} \ln \left[1-e^{-L_z \tilde{k}^{[l,s]}_z} \right], \label{Lifshitz_finiteT} \\ 
\tilde{k}^{[l,s]}_z =& \sqrt{\left(\sqrt{ m^2 + \left(2l+1-s \zeta \right)|qB|} - s \kappa q B \right)^2 + \xi_\mathcal{N}^2}. \notag
\end{align}

\begin{figure}[tb!]
    \centering
    \begin{minipage}[t]{1.0\columnwidth}
    \includegraphics[clip,width=1.0\columnwidth]{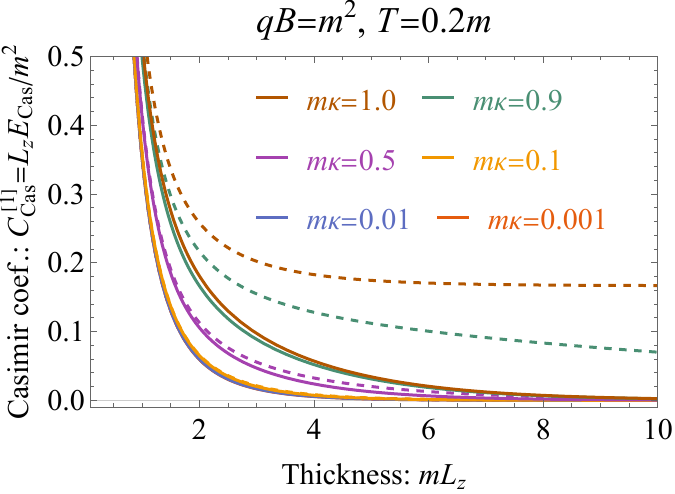}
    \end{minipage}
\caption{Casimir coefficient $C_\mathrm{Cas}^{[1]} \equiv L_z E_\mathrm{Cas}$ without ($\kappa=0$) and with AMM ($\kappa \neq 0$) at finite temperature $T=0.2m$.
The dashed lines are at $T=0$, which is the same as Fig.~\ref{fig:typ1}.}
\label{fig:finiteT}
\end{figure}

As a typical behavior at finite temperature, Fig.~\ref{fig:finiteT} shows the results at $T=0.2m$.
We find that the temperature effect suppresses the long $L_z$-behavior in the Casimir energy, which is a general feature in the case of the fermion fields (i.e., the effective gap effect due to the fermionic Matsubara frequency).
Therefore, the enhancement of the Casimir energy due to the AMM can be affected by finite temperature.

\subsection{At finite density}
Finally, we discuss a possible impact of AMM on the Casimir effect at finite fermion chemical potential or fermion density.
For studies about Dirac-fermion matter with AMM under magnetic field, see, e.g., Refs.~\cite{Chiu:1968zz,Strickland:2012vu,Ferrer:2015wca}.
The situation we consider can be realized when the size of materials consisting of fermions is extremely thin, e.g., in electron gas confined inside a thin cavity or a thin size of quark matter.
At finite chemical potential $\mu$, the dispersion relations are
\begin{align}
\omega_{l, s} &= E_{l, s} -\mu, \\
\tilde{\omega}_{l, s} &= -E_{l, s} -\mu,
\end{align}
where $E_{l, s}$ is the same as Eq.~(\ref{eq:E_ls}).
The corresponding Lifshitz formula is
\begin{align}
E_\mathrm{Cas} =&-2\int_{-\infty}^\infty \frac{d\xi}{2\pi} \sum_{l,s} \frac{|qB|}{2\pi} \ln \left[1-e^{-L_z \tilde{k}^{[l,s]}_z} \right], \label{Lifshitz_finitemu} \\ 
\tilde{k}^{[l,s]}_z =& \sqrt{\left(\sqrt{ m^2 + \left(2l+1-s \zeta \right)|qB|} - s \kappa q B \right)^2 - (i\xi + \mu)^2}. \notag 
\end{align}
This formula includes both the contributions from the Dirac-fermion vacuum and the finite-density environment.
When $\mu=0$, this formula reduces to Eq.~(\ref{Lifshitz}).

\begin{figure}[bt!]
    \centering
    \begin{minipage}[t]{1.0\columnwidth}
    \includegraphics[clip,width=1.0\columnwidth]{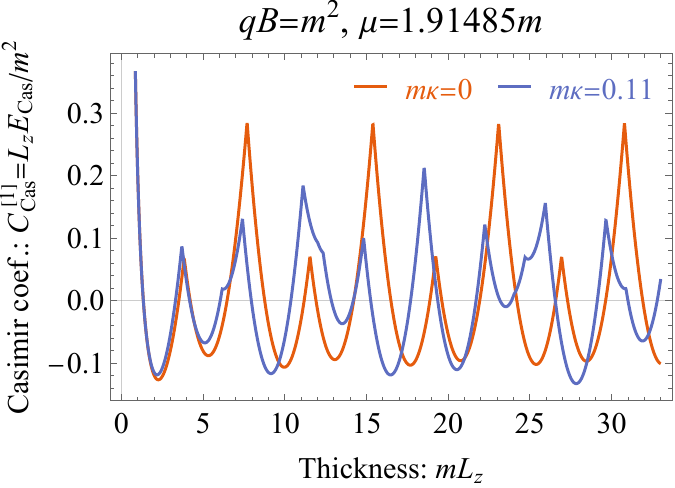}
    \end{minipage}
\caption{Casimir coefficient $C_\mathrm{Cas}^{[1]} \equiv L_z E_\mathrm{Cas}$ without ($\kappa=0$) and with AMM ($\kappa \neq 0$) at finite density $\mu=1.91485m$.}
\label{fig:finite-mu}
\end{figure}

In Fig.~\ref{fig:finite-mu}, we compare results at finite density with $\kappa=0$ and $\kappa \neq0$, where the parameters are adjusted such that the Fermi level crosses the three modes of $\omega_{l=0,+}$, $\omega_{l=0,-}$, and $\omega_{l=1,+}$.
It is known that when an energy level crosses the Fermi level, the Casimir energy oscillates as a function of $L_z$, and its period is determined by the Fermi momentum~\cite{Fujii:2024fzy,Fujii:2024ixq,Fujii:2024woy}.
At $\kappa=0$, the lowest state is the LLL $\omega_{l=0,+}$, and the second-lowest states, $\omega_{l=0,-}$ and $\omega_{l=1,+}$, are doubly degenerate.
Then, the Casimir energy oscillates, which is characterized by two oscillation periods, $mL_z^\mathrm{osc} \sim 3.85$ and $7.70$.
On the other hand, at $\kappa \neq 0$, $\omega_{l=0,-}$ and $\omega_{l=1,+}$ are split by the AMM.
Then, the oscillation of the Casimir energy is characterized by three oscillation periods.
Thus, a nonzero AMM can induce a new oscillation period by the level splitting between two modes.
In other words, the presence of multiple oscillation periods can be evidence of the AMM.

Note that in Fig.~\ref{fig:finite-mu}, we have focused on the situation with the three Landau levels, but in principle, this discussion holds even when more Landau levels cross the Fermi level.

\section{Discussion} \label{Sec:4}

\subsection{Other quantities}

In general, in thin materials or systems composed of fermions, the fermionic Casimir pressure can lead to expansion or compression of the material.
Such a deformation effect is a physically observable phenomenon.
The Casimir pressure is defined as the derivative of the Casimir energy with respect to thickness and can be straightforwardly calculated in our framework (e.g., see Ref.~\cite{Fujii:2024ixq}).

Furthermore, the fermionic Casimir effect influences various other physical quantities, such as specific heat (i.e., the temperature derivative), fermion density (the chemical-potential derivative), and magnetization or magnetic susceptibility (the magnetic-field derivatives).
Since Dirac fermions have a magnetic moment, intrinsic or dynamical AMMs will be useful for magnetically controlling these quantities.
In this sense, the study at finite temperature or density is significant.

\subsection{Numerical lattice simulations}

It is usually difficult to theoretically calculate Casimir effects in nonperturbative systems such as strongly interacting fermion systems, while one can employ numerical simulations of lattice quantum field theory.
In particular, the enhancement of the Casimir energy induced by the AMM will be quantitatively tested using future lattice QED, lattice NJL, or lattice QCD calculations (e.g., see Refs.~\cite{Chernodub:2018pmt,Chernodub:2018aix,Kitazawa:2019otp,Chernodub:2023dok} for the Casimir effect from Yang–Mills fields, which are a fundamental part of QCD).

\subsection{Quark matter}

In relativistic heavy-ion collision experiments, quark-gluon plasma (or nuclear matter) is produced as a fireball with a characteristic size of a few femtometers, which can be regarded as fermion matter with a thickness (see, e.g., Refs.~\cite{Fujii:2024fzy,Fujii:2024woy}).
Within this fireball, when a strong magnetic field is generated, the Casimir energy from the quark fields may be enhanced by the AMM of the quarks. Consequently, this enhancement could have a non-negligible impact on the fireball’s thermodynamics.

\subsection{Dirac or Weyl semimetals}

Thin films of Dirac or Weyl semimetals (see Refs.~\cite{Armitage:2017cjs,Lv:2021oam} for reviews) are other candidates of physical systems in which a fermionic Casimir effect occurs, where Dirac or Weyl quasifermion fields inside thin films induce a fermionic Casimir effect~\cite{Nakayama:2022fvh}.
The splitting of spin degeneracy in Dirac semimetals caused by a magnetic field, can be regarded as an AMM term in the effective Dirac equation.
With suitable modifications, our method can be extended to such systems.

\section{Conclusion} \label{Sec:5}
In this paper, we have discussed the impact of AMM on the Casimir effect originating from Dirac-fermion fields under a magnetic field.
From the main formula, Eq.~(\ref{Lifshitz}), we have found the following features:
\begin{enumerate}
\item {\it Casimir energy enhancement}---The Casimir energy in the long-$L_z$ region is enhanced by a nonzero AMM.
This is because the Casimir energy in this region is dominated by the LLLs, and the lower energy shift of the LLL increases the Casimir energy.
\item {\it Enhancement from HLLs}---The Casimir energy in the short-$L_z$ region is also enhanced by a nonzero AMM, where the Casimir energy is dominated by the sum of HLLs as well as the LLLs.
\item {\it Significant enhancement from gapless LLs}---When the AMM (and/or the magnetic field) is large enough, the Casimir energy significantly increases.
This is because the dispersion relations of the LLLs become an effectively gapless form.
\end{enumerate}
Thus, the enhancement of the Casimir energy is a robust property in many cases of the Dirac fields.
This feature will be useful for controlling the fermionic Casimir effect by magnetic fields.

\section*{ACKNOWLEDGMENTS}
The authors thank Mamiya Kawaguchi for fruitful discussions about quark AMMs.
This work was supported by the Japan Society for the Promotion of Science (JSPS) KAKENHI (Grants No. JP24K07034, JP24K17054, JP24K17059).

\section*{DATA AVAILABILITY}
The data that support the findings of this article are not publicly available. The data are available from the authors upon reasonable request. 

\appendix 

\section{Magnetic moments of constituent quarks} \label{App:quarkAMM}
In Subsection~\ref{subsec:quark}, the AMMs of constituent quarks were used as input parameters.
In this Appendix, we estimate the AMMs of constituent quarks based on the method proposed in Refs.~\cite{Bicudo:1998qb,Fayazbakhsh:2014mca}.
In our analysis, the most recent values in the Particle Data Group~\cite{ParticleDataGroup:2024cfk}  are used as input parameters.
The experimental values for the magnetic moments for the proton, neutron, and $\Lambda$ baryon are~\cite{ParticleDataGroup:2024cfk}
\begin{align}
\mu_p&= 2.7928473446 \, \mu_N, \\
\mu_n&= -1.9130427 \, \mu_N, \\
\mu_\Lambda&= -0.613 \, \mu_N,
\end{align}
respectively, where $\mu_N \equiv \frac{e}{2m_p}$ is the nuclear magneton with the elementary charge $e$ and the proton mass $m_p$.
In the constituent quark model~\cite{Franklin:1968pn}, the magnetic moments of the constituent quarks are expressed as
\begin{align}
\mu_u&=\frac{4\mu_p+\mu_n}{5} = 1.85167\,\mu_N, \\
\mu_d&= \frac{\mu_p+4\mu_n}{5} =-0.971865 \,\mu_N, \\
\mu_s&= \mu_\Lambda =-0.613 \,\mu_N.
\end{align}
Here, we define the magnetic moments of constituent quarks as $\mu_f =\frac{\bar{q}_fe}{2M_f} (1+a_f)$, where the quark flavor is labeled by $f=u,d,s$, $M_f$ is the constituent quark mass, and the magnitudes of electric charges are $\bar{q}_u=2/3$ and $\bar{q}_d=\bar{q}_s=-1/3$.
Then, we obtain the following expression for the constituent-quark AMM: 
\begin{align}
a_f = \frac{2M_f \mu_f}{\bar{q}_fe} -1 = \frac{M_f \mu_f}{\bar{q}_f m_p \mu_N} -1,
\end{align}
where we used $e = 2m_p \mu_N$.
We can find that, in the last expression, the unknown parameters are only $M_f$, since $\mu_f$, $\bar{q}_f$, and the proton mass $m_p=938.27208816$ MeV are already fixed by the experimental values.

For the $u$ and $d$ quarks, when we assume the constituent quark masses as $M_{u}=M_d = 340$ MeV,
\begin{align}
a_{u} &= 0.00647922 \to \kappa_{u} = 0.00952827 \, \mathrm{GeV}^{-1}, \label{eq:kappau}\\
a_{d} &= 0.0565187 \to \kappa_{d} = 0.0831157 \, \mathrm{GeV}^{-1}, \label{eq:kappad}
\end{align}
where we used $\kappa_f \equiv a_f/2M_f$.
These values are consistent with $\kappa_{u} = 0.00995 \, \mathrm{GeV}^{-1}$ and $\kappa_{d} = 0.07975  \, \mathrm{GeV}^{-1}$, which were estimated using $M_u=M_d = 340$ MeV as Eq.~(2.13) of Ref.~\cite{Fayazbakhsh:2014mca}.
We note that, as another estimate in Ref.~\cite{Fayazbakhsh:2014mca}, larger AMMs, $\kappa_u=0.29016 \, \mathrm{GeV}^{-1}$ and $\kappa_d= 0.35986 \, \mathrm{GeV}^{-1}$, were obtained in Eq.~(2.12) of Ref.~\cite{Fayazbakhsh:2014mca}.
However, its estimate is based on a somewhat larger constituent quark mass $M_u=M_d = 420$ MeV.
Therefore, we discard those values.

For the strange quark, when we assume $M_s = 550$ MeV,
\begin{align}
a_{s} &= 0.0779922 \to \kappa_{s} = 0.070902 \mathrm{GeV}^{-1}.
\end{align}
A similar value $a_s=0.053$ was obtained in Ref.~\cite{Aguirre:2020tiy}, where $M_s=538$ MeV was assumed.

\section{Dynamical quark mass in the NJL model} \label{App:NJL}

In Subsection~\ref{subsec:quark}, the masses of constituent quarks under magnetic fields were used as input parameters to quantitatively calculate the Casimir energy.
In this Appendix, we briefly explain how to estimate the magnetic-field dependence of dynamical quark masses with AMM.
In particular, we use the two-flavor NJL model~\cite{Nambu:1961tp,Nambu:1961fr}, which is a low-energy effective model with four-point coupling between quarks and which is useful for investigating the mass generation due to the spontaneous chiral symmetry breaking.
For many variations about quarks with an AMM under an external magnetic field, see Refs.~\cite{Fayazbakhsh:2014mca,Chaudhuri:2019lbw,Ghosh:2020xwp,Xu:2020yag,Mei:2020jzn,Aguirre:2020tiy,Wen:2021mgm,Chaudhuri:2021skc,Wang:2021dcy,Chaudhuri:2021lui,Lin:2022ied,Aguirre:2021ljk,Kawaguchi:2022dbq,Mao:2022dqn,Sheng:2022ssp,Chaudhuri:2022oru,He:2022inw,Qiu:2023kwv,Tavares:2023oln,He:2024gnh,Mondal:2024eaw,Kawaguchi:2024edu,Yang:2025zcs}.

\begin{figure}[b!]
    \centering
    \begin{minipage}[t]{1.0\columnwidth}
    \includegraphics[clip,width=1.0\columnwidth]{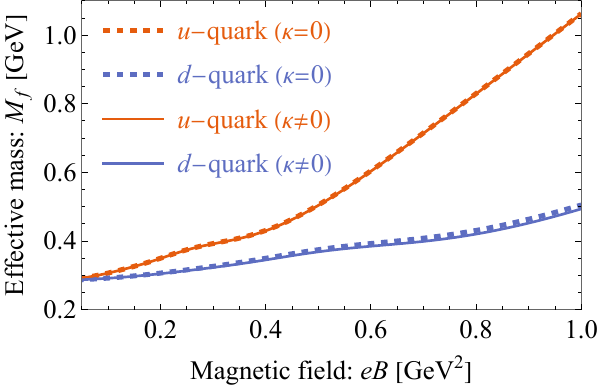}
    \end{minipage}
\caption{Magnetic-field dependence of effective masses of the up and down constituent quarks with $\kappa=0$ and $\kappa\neq 0$.}
\label{fig:NJL}
\end{figure}

In the NJL model, we assume the mean fields $(\sigma_u,\sigma_d)$ characterizing the chiral condensates for the up and down quarks and neglect the flavor-mixing effect, such as the $\sigma_u\sigma_d$ term, for simplicity.
The effective potential from the mean-field Lagrangian is written as
\begin{align}
\Omega_\mathrm{NJL} 
=& -N_c \sum_{f=u,d} \frac{|q_fB|}{2\pi} \sum_{s=\pm} \sum_{l=0}^\infty  \int \frac{dk_z}{2\pi} \omega_{l,s}^{[f]} F_\Lambda \notag\\
&+ \frac{1}{4G} \left(\frac{\sigma_u^2}{2} + \frac{\sigma_d^2}{2} \right), \label{eq:Omega_NJL}
\end{align}
where the notation is similar to that of Eq.~(\ref{E0int}).
$N_c=3$ is the number of colors, and $G$ is the four-point coupling constant between quarks.
The dispersion relations of constituent quarks with AMM $\kappa_f$ are
\begin{align}
\omega_{l,s}^{[f]} =& \sqrt{k_z^2 +  \left( \sqrt{M_f^2 +  |q_fB| \left(2l + 1 - s \zeta \right)} - s \kappa_f q_f B \right)^2}, 
\end{align}
where $M_f = m_0+\sigma_f$ is the effective mass with the current quark mass $m_0$.
For the regularization function $F_\Lambda$ with a cutoff parameter $\Lambda$, we adopt a magnetic-field-dependent Lorenzian-type function~\cite{Fukushima:2010fe,Gatto:2010qs,Gatto:2010pt},
\begin{align}
F_\Lambda =& \frac{\Lambda^{10}}{\Lambda^{10} + \sqrt{k_z^2 +|q_fB| (2l+1-s \zeta)}^{10}}.
\end{align}
For the model parameters, we use $\Lambda= 0.68138 \mathrm{GeV}, \, G = 1.860/\Lambda^2, \, m_0 = 4.552 \mathrm{MeV}$~\cite{Avancini:2019wed}.
By minimizing the effective potential (\ref{eq:Omega_NJL}) with respect to $(\sigma_u,\sigma_d)$, we can determine their values.

In Fig.~\ref{fig:NJL}, we show the numerical results for the effective masses, $M_u$ and $M_d$, where we used the AMMs in Eqs.~(\ref{eq:kappau}) and (\ref{eq:kappad}).
Thus, the mass shift for the up quark is more rapid than that for the down quark due to the difference between their electric charges.
The effect of the AMM is quite small (within the current parameters), but it contributes to a negative mass shift.

\bibliography{ref}

\end{document}